\documentclass[pra,twocolumn,amsmath,amssymb,aps]{revtex4-1}

\usepackage{graphicx}
\usepackage{amssymb}
\newcommand{\abbrev}[1]{\textsc{#1}}

\begin{document}

\title{Mesoscopic supersolid of dipoles in a trap}

\author{A. E. Golomedov$^1$, G. E. Astrakharchik$^2$, and Yu. E. Lozovik$^{1,3}$} 
\affiliation{
$^1$Institute of Spectroscopy, 142190 Troitsk, Moscow region, Russia\\
$^2$Departament de F\'{\i}sica i Enginyeria Nuclear, Campus Nord B4-B5, Universitat Polit\`ecnica de Catalunya, E-08034 Barcelona, Spain\\
$^3$Moscow Institute of Physics and Technology (State University), Dolgoprudny, Moscow Region, Russia 
}
\pacs{03.75.Hh, 05.30.Fk, 67.80.K-}

\begin{abstract}
A mesoscopic system of indirect dipolar bosons trapped by a harmonic potential is considered. The system has a number of physical realizations including dipole excitons, atoms with large dipolar moment, polar molecules,  Rydberg atoms in inhomogenious electric field. We carry out a diffusion Monte Carlo simulation to define the quantum properties of a two-dimensional system of trapped dipoles at zero temperature. In dimensionless units the system is described by two control parameters, namely the number of particles and the strength of the interparticle interaction. We have shown that when the interparticle interaction is strong enough a mesoscopic crystal is formed. As the strength of interactions is decreased a multi-stage melting takes place. Off-diagonal order in the system is tested using natural orbitals analysis. We have found that the system might be Bose-condensed even in the case of strong interparticle interactions. There is a set of parameters for which a spatially ordered structure is formed while simultaneously the fraction of Bose condensed particles is non zero. This might be considered as a realization of a mesoscopic supersolid.
\end{abstract}

\maketitle

\section{Introduction}
Bose-Einstein condensation (BEC) is a phenomenon that has been attracting great attention for a long time since its prediction in 1924 \cite{Einstein1,Einstein2}. Theoretical description of condensate properties is commonly based on mean-field Gross-Pitaevskii equation \cite{PitaevskiiBECbook} which applies to a dilute Bose gas. This assumption is not fulfilled in systems with strong correlations.

This is the case of a two-dimensional rather dense dipole system of indirect excitons in coupled quantum wells or in single quantum well in strong electric field \cite{lozEx}. At the densities much smaller than $a^{-2}$ (with $a$ being the characteristic excitons size) exchange effects are greatly suppressed by the dipole-dipole repulsion, so dipole excitons can be treated as Bose particles. The exciton confinement can be created by inhomogeneous  electric field or inhomogeneous deformation of semiconductor. The inhomogeneous electric field can be generated, for example, by a tip of scanning probe-microscope or by a profiled controlling gate (see \cite{timofeev:081708} and references therein).

The dipolar interactions are also important in atomic gases with large dipolar moment. The Bose-Einstein condensation has been achieved in $^{52}$Cr atoms possessing large permanent dipolar moment \cite{PhysRevLett.94.160401}. In this system there is a competition between short-range interactions and long-range dipolar forces. Strong dipolar effects has been experimentally observed in trapped chromium atoms, see Ref. \cite{PhysRevLett.95.150406, PhysRevLett.97.250402, NatureCr52, NatureDipoles}. Large dipolar moment can be induced in polar molecules by applying an external electric field. Recently the quantum regime of $^{40}$K$^{87}$Rb fermionic polar molecules has been reached and dipolar collisions in such systems has been experimentally studied \cite{Ospelkaus12022010, NaturePolarMolecules}. Rydberg atom (atom with one electron excited to a very high principal quantum number \cite{RevModPhys.82.2313}) has a very large size and a very large polarizability, i.e. large dipole moment can be inducted in moderate electric field. An electric field align dipolar moments so that the interaction in $2$D system has dipolar $1/r^3$ form. An optical trapping can be used to confine the system of Rydberg atoms \cite{PhysRevLett.104.223002}.\\

A close relation between two quantum phenomena, BEC and superfluidity, suggests that a system might be superfluid if a part of it is Bose condensed. Moreover,  at the same time the system can possess a crystal order, i.e. can be a supersolid. A finite superfluid signal was reported in a number of recent experiments with cold helium in solids (see \cite{E.Kim09242004,PhysRevLett.97.115302, PhysRevLett.98.175302,springerlink:10.1007/s10909-007-9471-1, springerlink:10.1007/s10909-007-9452-4,PhysRevLett.99.015301, 1742-6596-150-3-032040, Hunt01052009}). It is probable that the superfluid signal observed in experiments is due to presence of defects. We note that mesoscopic trapped systems are good candidates for being supersolid as no perfect commensurate crystalline structure can be formed and defects are intrinsically present in the system. We perform numerical simulations of a trapped system of excitons with dipolar interaction, to obtain structure  properties and to study how the crystal-like order is formed in the system.

In the regime when dipole-dipole repulsions are weak, analytical approaches based on mean field approximation, provide a good description of the system, but fail when the density is large. Instead Monte Carlo methods have no such limitations and can be successfully applied to strongly interacting systems. In Ref.~\cite{BelousovCMC} properties of a classical dipolar system in a harmonic trap at low finite temperature were studied. Path Integral Monte Carlo (PIMC) simulation of trapped clusters were performed in works of Lozovik et al. \cite{JETP.79.10.473} and Pupillo et al. \cite{PhysRevLett.104.223002} for dipolar interaction and for Coulomb interaction by Filinov et al. \cite{PhysRevB.77.214527}. PIMC technique is useful for simulation of the properties of quantum systems at a finite temperature, while the smaller is the temperature the more difficult it is to obtain accurate results, which makes the simulation of the ground state of the system extremely difficult. Instead diffusion Monte Carlo method is well suited for studying ground state properties. In references \cite{AstraBook, PhysRevLett.98.060405, PhysRevLett.98.060404,PhysRevB.82.014508} two-dimensional (2D) systems of dipoles in the absence of an external potential were studied. It was shown that at large density a crystal is formed. Non-zero superfluid and condensate fractions have been found in finite-size crystals and in crystals with vacancies.

In the present work we study effects of a harmonic confinement in a 2D geometry. Our motivation for expecting a  coexistence of a finite condensate fraction and broken rotational symmetry (i.e. ``supersolidity'') in trapped systems are two-fold as: (i) small trapped systems are mesoscopic (ii) inherent incommensurability between spherical geometry of the trap and triangular geometry of a 2D triangular lattice leads to an effective introduction of defects in the lattice, which might lead to appearance of superfluidity. To our best knowledge quantum  systems of two-dimensional trapped dipoles at zero temperature so far were not well studied.

This work is organized as follows. In Section~\ref{sec:model} we outline model Hamiltonian and describe methods used in simulation, in Section~\ref{sec:BEC} we describe method of natural orbitals used for studying coherence in the system in this section we generalized method used in 3D system (see \cite{PhysRevA.63.023602}, \cite{PhysRevB.37.4950}) to the case of 2D, in Section~\ref{sec:Results} we present obtained results and in Section~\ref{sec:sonclusion} we draw conclusions.

\section{Model system and numerical approach}\label{sec:model}
We consider a two-dimensional system of $N$ dipolar bosons in a harmonic trap with frequency $\omega$. Such a system is described by the
Hamiltonian
\begin{equation}
\hat{H} = \sum_i^N -\frac{\hbar^2}{2 m} \nabla^2_i +\sum_i^N \frac{1}{2} m
\omega^2 r_i^2 +\sum_{i<j}^N \frac{d_{dip}^2}{r_{ij}^3}.
\label{Eqn:Hamiltonian}
\end{equation}
The second term in the Hamiltonian~(\ref{Eqn:Hamiltonian}) is associated with harmonic confinement potential, the third term describes the dipolar interaction between particles, $m$ is the mass of a particle and $d_{dip}$ is its dipole moment. It is convenient to use oscillator units in the problem, that is to measure length in units of oscillator length $a_0 = \sqrt{\hbar/m \omega}$ and energy in units of $\hbar \omega$. The dimensionless Hamiltonian is then
\begin{equation}
	\hat{H} = \frac{1}{2} \sum_i^N \left(-\nabla^2_i +
	r_i^2\right)+\sum_{i<j}^N \frac{d}{r_{ij}^3},
\end{equation}
where $d = d_{dip}^2/a_0^3 \hbar \omega$ is a dimensionless coupling parameter, which can be interpreted as the ratio of the typical energy of the dipolar interaction energy $E_{int} = d_{dip}^2/a_0^3$ and the characteristic energy of a harmonic oscillator confinement $E_{trap} = \hbar \omega$.

In order to study system properties we use Monte Carlo (MC) methods. A number of MC methods may be employed for simulation of quantum systems: variational Monte Carlo (VMC), diffusion MC, Path Integral Monte Carlo (PIMC), Path Integral Ground State etc. In the present work we are interested in ground state (zero temperature) properties and we do so by means of VMC and DMC techniques.

In VMC method one samples particle distribution according to a chosen trial wave function. By doing that it is possible to obtain mean values of an observable by averaging its value over the chain of realizations of particles coordinates. Method proposed by Metropolis et al. \cite{metropolis:1087} is used to generate such a chain.

Although we do not know the many-body wave function exactly, some physically sound ansatz may be used for trial wave function so that it depends on particles coordinates and some a set of additional parameters. Those parameters, referred to as variational parameters, are used to minimize energy corresponding to this wave function. A variational principle applies, that is the variational energy calculated in this way is always larger than the ground state energy and equal to it when the trial wave function coincides with the ground state wave function

The main idea of the DMC technique is to solve the Schr\"odinger equation in the imaginary time. For sufficiently long evolution of an arbitrary wave function in the imaginary time its projection to the excited states is exponentially suppressed compared to its projection to the ground state. The problem is that one would rather prefer to sample the square of the absolute value of the wave function (probability density) than the wave function itself, because average values of an observable with diagonal operator is defined as an integral where observable is integrated with weight equal to absolute value of the wave function squared. But it is impossible to introduce a closed real-valued equation defining the time evolution of the probability density. A common choice is to calculate mixed estimators for observables where observable is integrated with weight equal to $f (\mathbf{R}, t) = \psi_T(\mathbf{R}) \psi_0(\mathbf{R}, t)$, with $\psi_T(\mathbf{R})$ being the trial wave function and $\psi_0(\mathbf{R}, t)$ the ground state wave function. A mixed estimator introduces some bias due to a particular choice of the trial wave function, unless the observable commutes with the Hamiltonian. An important example of a mixed estimator being exact is in the evaluation of the ground state energy.

For other observables one can reduce this bias by extrapolating mixed estimator to the exact one,
\begin{equation}
	A_{extr} = 2 A_{mixed} - A_{trial}.
\end{equation}
Here $A_{mixed} = \int \psi_T A \psi_0$ is mixed estimator of an observable $A$ and $A_{trial} = \int \psi_T^{*} A \psi_T$ is a variational estimator. This extrapolation is accurate to the second order of difference between trial wave function and exact one. We applied the extrapolation for all observables that not commute with Hamiltonian.

To reduce the extrapolation error it is important that the projection of the trial wave function to the ground state is as large as possible. We construct the trial wave function in the Nosanow-Jastrow form
\begin{equation}
\Psi_T(\mathbf{R}) = \prod_i^N f(r_i) \prod_{i<j}^N g(r_{ij}), \label{Eqn:TWF_Basic}
\end{equation}
where $\mathbf{R} = \{r_1,r_2, \ldots r_N\}$ stands for a point in $2N$-dimensional phase space, $f(r)$ and $g(r)$ are one- and two-body correlation terms. The first term in $\Psi_T(\mathbf{R})$ takes into account one-body physics and describes the effects of the harmonic oscillator, while the second term introduces interparticle correlations. For different values of the interaction strength $d$ we have used several forms of $f(r)$ and $g(r)$.

\subsection{Gas of dipoles}
An exact expression for the one-body terms $f(r_i)$ of the wave function~(\ref{Eqn:TWF_Basic})is known in the limit of an ideal Bose gas $d  \rightarrow 0$ and is given by Gaussians $f_{ho}(r_i)=\exp{(-r_i^2/2a_0^2)}$. We keep this functional dependence form for finite values of the interaction strength $d$:
\begin{equation}
f(r) = \exp{(-\alpha r^2)},
\end{equation}
with a single variational parameter $\alpha$, which value is fixed by minimizing the variational energy.

The large distance physics are dominated by the Gaussian dependence in $f(r)$, so the main requirement for the two-body Jastrow term is to describe correctly short-range physics. When two particles come close to each other the influence of other particles can be neglected and the two body Jastrow term in the trial wave function~(\ref{Eqn:TWF_Basic}) can be well approximated by the zero-energy scattering solution of the two-body problem,
\begin{equation}
g(r) = K_0\left(2 \sqrt{d/r}\right).
\end{equation}
The resulting trial wave function is,
\begin{equation}
	\psi(\mathbf{R}) = \prod_i^N \exp{(-\alpha r_i^2)} \prod_{i\neq j}^N K_0\left(2 \sqrt{d/r_{ij}}\right). \label{Eqn:TWF_Gas_Final}
\end{equation}
\subsection{Crystal of dipoles}
When a crystal is formed the system loses the translational symmetry. To take this into account one has to use a proper symmetry in the trial wave function. We introduce a localizing ``crystal'' term in TWF,
\begin{equation}
 u(\mathbf{R}) = \prod_i^{N_c} \sum_i^N \exp{(-\beta
 (\mathbf{r_i}-\mathbf{r_j^c})^2)}, \label{TWF_crystal}
\end{equation}
where $\beta$ is a variational parameter. Each Gaussian term describes particle localization close to sites $r_j^c$, the total number of sites in crystal being denoted by $N_c$. We have used classical Monte Carlo method combined with gradient descent optimization to obtain position of sites $r_j^c$ for given particle number $N_c$. After obtaining them we have spread classical coordinates with a factor $\gamma$ optimized by means of variational Monte Carlo.

Combining Eq.~(\ref{TWF_crystal}) with trial wave function of gas Eq.~(\ref{Eqn:TWF_Gas_Final}) we obtain the final expression for the crystal trial wave function:
\begin{eqnarray}
\psi(\mathbf{R}) =  \left( \prod_i^N \exp{(-\alpha r_i^2)} \prod_{i\neq j}^N
K_0\left(2 \sqrt{d/r_{ij}}\right) \right) \nonumber \\
\times \prod_i^{N_c} \sum_i^N \exp{(-\beta
(\mathbf{r_i}-\mathbf{r_j^c})^2)}.
\end{eqnarray}

\begin{figure}
\includegraphics[width=\columnwidth,angle=0]{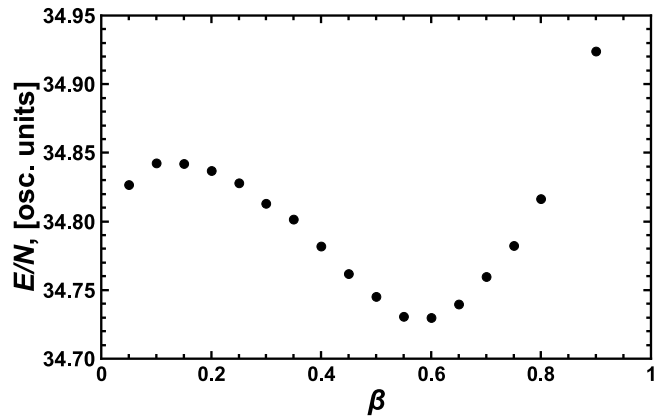}
 \caption{Variational energy as a function of the variational parameter $\beta$
 in a cluster of $N=32$ particles with interaction strength $d=100$.}
 \label{fig_cryst_minima}
\end{figure}

The typical dependence of the variational energy on parameter $\beta$ is shown in Fig.\ref{fig_cryst_minima}. There are two minima. The first one at $\beta=0$, corresponds to delocalized system or a gas state. The second minimum is located at $\beta\neq0$ and corresponds to a localized system or a crystal.

\subsection{Gas of strongly correlated dipoles}
Crystal and gas wave functions provide a good description in corresponding limits, but their use for intermediate values of the interaction strength produce large error in mixed estimators. The smaller is difference between the trial and ground state and wave functions, the smaller is the error. To reduce this difference in region of intermediate interaction strength, we introduce a shell term in the trial wave function,
\begin{equation}
	s(r) = \left[ \sum_{i=1}^{N_p} a_i \exp{(-(r-r_i^s)^2/\sigma_i^2)}
	\right]^\gamma,
\end{equation}
where $r_i^s$ is the separation of the $i$-th shell from the center of trap, $\sigma_i$ is the width of $i$-th shell and $a_i$ is its amplitude, $\gamma$ is a variational parameter. Parameters of shells are optimized so that $s(r)$ (for $\gamma = 1$) is the best approximation of the ratio of variational and diffusion radial distributions. Variational result for the radial distribution calculated with $s(r)$ reproduces closely DMC mixed estimator calculated without $s(r)$, and it means that the trial wave function is better when shell term $s(r)$ is used. The resulting trial wave function for this case is
\begin{widetext}
\begin{eqnarray}
\psi(\mathbf{R}) =
\prod_i^N \left(\exp{(-\alpha r_i^2)} \left[
\sum_{i=1}^{N_p} a_i \exp{(-(r-r_i^s)^2/\sigma_i^2)} \right]^\gamma\right)
\prod_{i\neq j}^N K_0\left(2 \sqrt{d/r}\right). \label{TWF:shell_structured}
\end{eqnarray}
\end{widetext}

\section{Bose-Einstein condensate}\label{sec:BEC}
To test the coherence in the system and to calculate the condensate fraction we study the spectral properties one-body density matrix (OBDM). The OBDM is defined as
\begin{equation}
\rho(\mathbf{r}, \mathbf{r^\prime}) = \left \langle
\hat{\Psi}^\dagger(\mathbf{r^\prime}) \hat{\Psi}(\mathbf{r})
\right
\rangle, \label{rho_basic}
\end{equation}
where $\hat{\Psi}(\mathbf{r})$ is the field operator that annihilates a particle at the point $\mathbf{r}$. Using single particle states $\phi_i(\mathbf{r})$, we expand the field operator,
\begin{equation}
\hat{\Psi}(\mathbf{r}) = \sum_i \phi_i(\mathbf{r}) \hat a_i,
\label{psi_expand_vector}
\end{equation}
where $\hat a_i$ is the bosonic annihilation operator that annihilates a particle in the state $\left \vert i \right \rangle$. Substituting
(\ref{psi_expand_vector}) into (\ref{rho_basic}) one obtains the spectral decomposition
\begin{equation}
\rho(\mathbf{r}, \mathbf{r^\prime}) = \sum_{ij} \phi_i^*(\mathbf{r})
\phi_j(\mathbf{r^\prime}) N_i \delta_{ij}. \label{rho_eig}
\end{equation}
The OBDM is diagonal in the single particle state basis, and natural orbitals $\phi_i(\mathbf{r})$ (occupation numbers $N_i$) are its eigenvectors (eigenvalues). The condensate is described by the orbital $\phi_0(\mathbf{r})$ with the macroscopic occupation number and condensate fraction is $n_0 = N_0/N$.

The definition of the OBDM operates with eigenfunctions $\phi(\mathbf{r})$ which are functions of four parameters in a 2D system, so that OBDM cannot be diagonalized using standard matrix methods. However the problem for the eigenvalues can be simplified if there is a cylindrical symmetry of the problem. In this case the OBDM depends on angle $\theta = \phi-\phi^{\prime}$ between vectors $\mathbf{r}$ and $\mathbf{r^\prime}$, rather than on two angles $\phi$ and $\phi^{\prime}$. Furthermore OBDM can be expanded in a series of angular momentum components:
\begin{equation}
\rho(\mathbf{r}, \mathbf{r^\prime}) = \frac{1}{\sqrt{2 \pi}} \sum_l
\rho_l(r,r^\prime) \exp{(il\theta)}
\end{equation}
with the projection of the OBDM to the state with angular momentum $l$, $l=0,
\pm 1 \ldots$ given by
\begin{eqnarray}
\rho_l(r,r^\prime) = \int d \theta d r_2 \ldots d r_N
\Psi^*(r,r_2,\ldots,r_N) \nonumber \\
\times \exp{(-il\theta)} \Psi(r,r_2,\ldots,r_N).
 \label{rho_l_calc}
\end{eqnarray}
Further, the one-body orbitals can be expanded in terms of the angular momentum
\begin{equation}
\begin{array}{c}
\phi_k(\mathbf{r}) = R_{nl}(r) \Phi_l(\varphi),\\
\Phi_l(\varphi) = \frac{1}{\sqrt{2\pi}} \exp{(il\varphi)},
\end{array} \label{eig_phi}
\end{equation}
where compound index $k$ consists of two indexes ${l,n}$. Substituting Eq.~(\ref{eig_phi}) into Eq.~(\ref{rho_eig}) we obtain the representation of the $l$-th OBDM in terms of natural orbitals
\begin{equation}
\rho_l(r,r^\prime) = \sum_i \phi_{il}^*(r) \phi_{il}(r^\prime) N_{il}.
\label{rho_l_eig}
\end{equation}
Matrices $\rho_l(r,r^\prime)$ can be sampled in Monte Carlo simulation according to Eq.~(\ref{rho_l_calc}) and are ``usual'' algebraic matrices understood as functions of two scalar arguments, which can be readily diagonalized using standard matrix methods. For condensate study we consider only components with angular moment $l=0$. To solve this equation numerically we introduce regularized functions
\begin{equation}
u_{il}(r) = \phi_{il}(r) \sqrt{r}.
\end{equation}
These functions $u_{il}(r)$ are better than $\phi_{il}(r)$ in calculations, because they are well behaved near $r=0$. In terms of $u_{il}$(r), the relation Eq.~(\ref{rho_l_eig}) reads as
\begin{equation}
\sqrt{r} \rho_l(r,r^\prime) \sqrt{r^\prime} = \sum_{il}u_{il}^*(r)
u_{il}(r^\prime) N_{il}.
\end{equation}
We solve this equation and obtain the condensate wave function
\begin{equation}
\phi_0(r) = u_{00}(r)/\sqrt{r}.
\end{equation}

\section{Results} \label{sec:Results}
In this section we present results of DMC and VMC simulations of the system. In the first part of the section we discuss structural properties of the system and in the second part we study properties of the condensate.

\subsection{Structural properties}
We perform a Monte Carlo study of structural and energetic properties of dipolar clusters by doing calculations with different trial wave functions. We have tested the quality of different trial wave functions for a cluster of $N=32$ particles in the strongly interacting regime $d=200$. The results obtained for the energy are summarized in Table~\ref{table_en}.
\begin{center}
\begin{table}[h!b!p!]
	\caption{Ground state energy per particle for $N=32$ particles for $d=200$, energy is measured in units of $\hbar \omega$. Number in parenthesis show the error on the last digit.}
		\centering
		\begin{ruledtabular}
		\begin{tabular}{ll}
		\textrm{Method and trial wave function}&
		\textrm{Energy per particle}\\
		\colrule
		VMC, gas TWF & 44.9584(9)\\
		DMC, gas TWF & 44.4201(6)\\
		VMC, crystal TWF & 44.848(1)\\
		DMC, crystal TWF & 44.4491(5)\\
		VMC, TWF with shell term & 44.8523(7)\\
		DMC, TWF with shell term & 44.4145(8)\\	
		\end{tabular}
		\end{ruledtabular}
	\label{table_en}
\end{table}
\end{center}
One can see that the best variational energy of the cluster with strong interactions $d=200$ is the one obtained using the crystal trial wave function~(\ref{TWF_crystal}). Energy of a crystal in DMC calculation is larger than both DMC energy of a gas and DMC energy of a strongly correlated gas with formed shells. The best description of the system is given by the wave function shell structured gas~(\ref{TWF:shell_structured}).

\begin{figure}
\includegraphics[width=\columnwidth,angle=0]{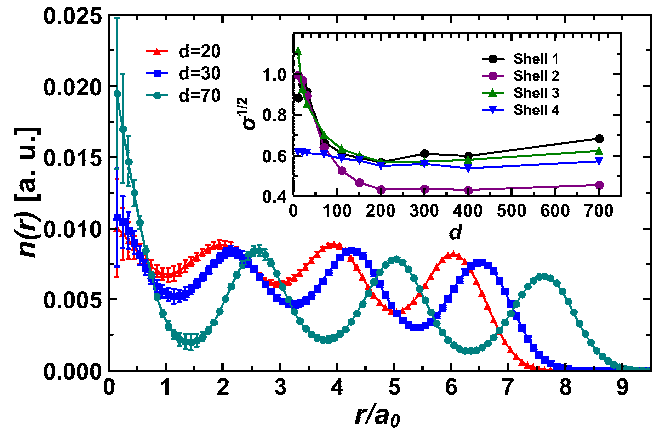}
 \caption{Radial particles distribution and width of the shells (inset) for
 different values of interaction strength $d$}
 \label{fig_dispersions}
\end{figure}

DMC calculations of the energy of the cluster with different interaction strengths showed that the gas phase is energetically preferable for values of interaction strength smaller than $\sim 100$, in the region of $100 \lesssim d \lesssim 400$ strong correlated gas is energetically preferable and for values of $d$ larger than $\sim 400$ a full crystal is formed. Figure~\ref{fig_dispersions} shows radial distribution of particles $R(r)$ for different values of $d$ obtained with gas trial wave function. One finds that $R(r)$ becomes nonmonotonic as the interaction increases and a shell structure is formed. We find that the shells can be well approximated by a set of Gaussians. The width of the Gaussians (see inset in Fig. \ref{fig_dispersions}) saturates to a constant value when $d$ is large enough. This fact can be explained if we assume that dipoles are moving around crystal sites in the mean field associated with other particles and consider a generalized interaction law $1/r^\alpha$, where $\alpha>1$. In a homogeneous crystal the first not disappearing term of the mean field is quadratic one with frequency  $\omega_{mf} \sim V_{real}^{\prime\prime}(r) \bigg|_{r=r_0} = \alpha (\alpha - 1) V(r_0) / r_0^2,$ where $r_0$ is the mean interparticle distance. In the presence of the external confinement the size of the cloud is proportional to the interparticle distance $r_0$ and is such that characteristic trapping energy is of the same order as the interparticle energy, that is $V(r_0) \sim \omega_{trap} r^2.$ Taking this into account, one finds that particle displacement in mean field $\Delta$ is constant for given trap frequency,
$\Delta \sim \alpha (\alpha - 1)\omega_{trap}, $ with $\alpha=3$ for dipoles.
Results of VMC and DMC calculations for radial distribution differ for large values of $d\sim100$, which means that the gas trial wave function provides an inadequate description of the system in the regime of strong interactions. We repeated our calculation for trial wave function with shell terms and obtained that radial ordering decreases the energy, but the lowest energy corresponds to both radial and angular localization, i.e. to a crystal phase.

\begin{figure}
\centering
\includegraphics[width=\columnwidth,angle=0]{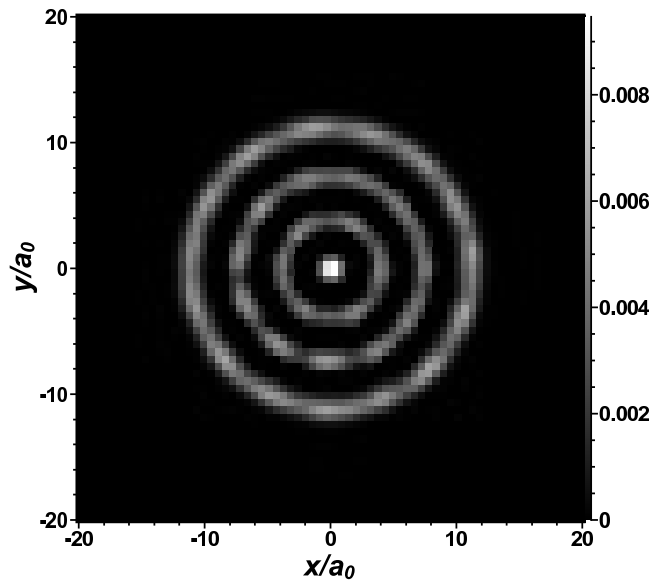}
 \caption{($x$,$y$) density profile for $N=32$ particles in the trap for
 the interaction strength $d=500$}
 \label{fig_2ddistr}
\end{figure}

\begin{figure}
\includegraphics[width=0.8\columnwidth,angle=0]{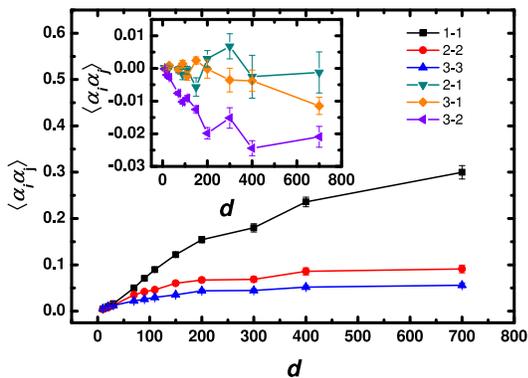}
 \caption{Diagonal (main graph)
 and nondiagonal (inset) components of the orientational order parameter $\left
 \langle \alpha_i \alpha_j \right \rangle$ with $\alpha_i$ defined as
 Eq~\ref{eqn_op} for cluster with $32$ particles}
  \label{fig_op}
\end{figure}

A contour plot of the density profile of particles is shown in Fig.~\ref{fig_2ddistr}. Darker color corresponds to lower density of particles. The internal structure of the shells is not resolved in Figure~\ref{fig_2ddistr}, as the system can freely rotate due to rotational symmetry. In order to extract information on the shells ordering we introduce inter- and intra- shell order parameter $\left \langle \alpha_i \alpha_j \right \rangle$
with values $\alpha_i$ defined according to
\begin{equation}
\alpha_j = \frac{1}{N_j} \sum_k \exp{(i N_j \phi_k)},
\label{eqn_op}
\end{equation}
where $\phi_k$ is the angle of $k-th$ particle in the $j-th$ shell containing $N_j$ particles. For $i=j$ the function $\left \langle \alpha_i \alpha_j \right \rangle$ is intra-shell order parameter and the larger is its value the stronger are internal correlations inside the shell. For $i\neq j$ the function $\left \langle \alpha_i \alpha_j \right \rangle$ is inter-shell order parameter and it measures the correlations strength between shells $i$ and $j$. Results of DMC simulation for a cluster of 32 particles with a gas trial wave function~(\ref{Eqn:TWF_Gas_Final}) are presented in Fig.~\ref{fig_op}. One finds that the intra-shell order parameter is significantly different from zero for interaction strength larger than $d\sim100$ and that the inter-shell order parameter is significantly different from zero only for shells 2 and 3 for interaction strength larger than $d\sim300$. Since we have used gas trial wave function this correlations are not caused by the symmetry of the trial wave function.

To summarize, we suggest the following scenario of $T=0$ quantum crystallization. When $d$ is small the gas state is energetically favorable, starting from $d\sim10$ shells are formed in the system, then up to $d\sim300$ shells solidify and become ordered inside. From $d\sim200$ shells start to order between each other and around $d\sim400$ a complete crystal is formed.

\subsection{Bose-Einstein Condensation and mesoscopic supersolid}

We applied the technique of natural orbitals described in Section \ref{sec:BEC}, to obtain the fraction of Bose-condensed particles in a cluster of $N=32$ particles with the interaction strength varying from infinitely small ($d=0$) to intermediately strong ($d\sim100$) values.

\begin{figure}
\includegraphics[width=0.8\columnwidth]{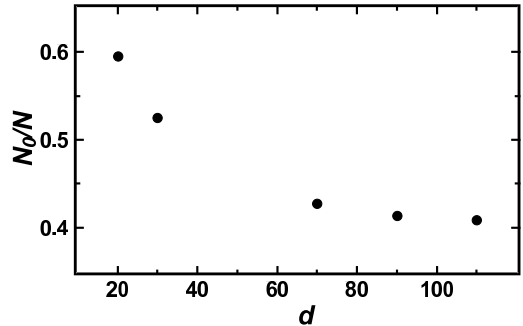}
 \caption{Fraction of condensed particles $N_0/N$ as a function of the interaction strength $d$, number of particles in cluster $N=32$}
  \label{fig:cond}
\end{figure}

To test the method used to estimate the condensate fraction we checked that for noninteracting system all particles are condensed. At finite values of the interaction strength, the condensate is suppressed. The dependence of the condensate fraction on the interaction strength is presented in Fig.~\ref{fig:cond}. Our calculations show that even in region of intermediate-strong interactions $d\sim100$ the fraction of condensed particles significantly differ from zero and is order of $40\%$. At the same time shell structure is already  formed. Simultaneous coexistence of broken rotational symmetry (shells) and a Bose-condensate can be referred to as a mesoscopic supersolid.

\section{Conclusions}
\label{sec:sonclusion}	
In the present work a two-dimensional system of bosonic dipoles in a harmonic trap is studied. Dipoles are assumed to be oriented perpendicularly to the two dimensional plane and to interact according to repulsive dipole-dipole potential. We found that cluster undergoes a quantum crystallization for large value of dipole interaction strength ($d \sim 400$). In the process of zero temperature transition interaction strength plays the role of a control parameter. We found that crystallization consists of three stages: formation of shells, in-shell ordering and ordering between shells.

We found that quantum Bose-Einstein condensation occurs in the system. We investigate dependence of the fraction of condensed particles on interaction strength up to the intermediate-strong interaction strengths, and find that number of particles in condensate can be sufficiently different from zero. In this regime spatial ordering (shells structure) coexists with Bose condensation, thus a mesoscopic supersolid is formed.

\acknowledgements
The work was sponsored by RFDR grant and RAS program. GEA acknowledges fellowship by MEC (Spain) and financial support by (Spain) Grant No.~\abbrev{fis}2008-04403, Generalitat de Catalunya Grant No.~2009\abbrev{sgr}-1003.


%

\end{document}